# HARMONIZATION AND EVALUATION
## TWEAKING THE PARAMETERS ON HUMAN LISTENERS

Filippo CARNOVALINI[a], Alessandro PELIZZO[a], Antonio RODÀ[a], Sergio CANAZZA[a]

[a] *Centro di Sonologia Computazionale (CSC), Dept. of Information Engineering, University of Padova, {filippo.carnovalini,antonio.roda,sergio.canazza}@unipd.it*

**ABSTRACT**

Kansei models were used to study the connotative meaning of music. In multimedia and mixed reality, automatically generated melodies are increasingly being used. It is important to consider whether and what feelings are communicated by this music. Evaluation of computer-generated melodies is not a trivial task. Considered the difficulty of defining useful quantitative metrics of the quality of a generated musical piece, researchers often resort to human evaluation. In these evaluations, often the judges are required to evaluate a set of generated pieces along with some benchmark pieces. The latter are often composed by humans. While this kind of evaluation is relatively common, it is known that care should be taken when designing the experiment, as humans can be influenced by a variety of factors. In this paper, we examine the impact of the presence of harmony in audio files that judges must evaluate, to see whether having an accompaniment can change the evaluation of generated melodies. To do so, we generate melodies with two different algorithms and harmonize them with an automatic tool that we designed for this experiment, and ask more than sixty participants to evaluate the melodies. By using statistical analyses, we show harmonization does impact the evaluation process, by emphasizing the differences among judgements.

## 1  INTRODUCTION

The capability of music to arouse various categories of feelings such as colors, feelings, or emotions (e.g., Juslin and Sloboda, 2011; Rodà et al., 2014; Murari et al., 2015; Rodà et al., 2018) has many applications. In the information technology field, a musical signal can contribute to the multimodal/multisensory interaction, providing the user with information through sonification. In this sense, sound design requires great attention and a deep understanding of the influence of musical parameters on the user's experience. In this field, the use of automatically generated melodies is very interesting, to have a large selection of inexpensive custom music (necessary to provide sonification to many events with different characteristics). Unfortunately, despite the large number of studies on the connotative meaning of music, very few are related to the repertoire of music automatically generated. This work introduces this study for generated melodies, investigating the effect of harmony.

There is an abundance of methods in which music can be programmatically generated, as the scientific literature on the subject testifies (Herremans et al., 2017; Tatarand and Pasquier, 2019; Briot and Pachet, 2020; Carnovalini and Rodà, 2020).

Both artists and AI practitioners are interested in finding novel ways to generate music, for a variety of goals, and this has led to the design of many algorithms that are capable, to some extent, of generating musical material (Lamb, 2018). Most of these systems are focused on the generation of melodies, sometimes with the aim to imitate a particular style or composer.

This abundance of research on music generation is not as widely accompanied by research on how to evaluate if the quality of the generated material is satisfactory (Jordanous, 2013). While the most obvious way to assess this is referring to a musician or a musicologist who is expert on the kind of music that is being generated, often listening surveys asking to evaluate the musical outputs of an algorithm are used as a proxy of these expert evaluations. Such approach is often inspired by the Turing Test, although it should be noted that the Imitation Game envisioned by Turing does not really apply to this kind of evaluation (Ariza, 2009).

Despite others have underlined the risks of using such tests, since human listeners (especially if not experts (Soldier, 2002)) are not completely reliable and are subject to priming effects (Oore et al., 2018), to the best of our knowledge there is currently no study that tries to quantify just how susceptible humans are to the changes in how music is presented to them at the aural level. In other words, it is not completely clear if presenting two melodies in two different settings (such as a different arrangement) could change the preference between the two melodies expressed by the listener.

In this paper we propose a between-subjects study to evaluate if changing the aural setting (in this case, by harmonizing and arranging the melodies that need to be evaluated) has a significant effect on the evaluation of two algorithms for the generation of melodies. It is worth noting that while we choose two algorithms from scientific literature, the same process could simulate a test between generated tunes and human-composed ones.

## 1.1 Hypothesis

The main claim we wish to make and to prove is the following.

When presenting a set of computer-generated melodies in aural form to a human listener and asking them to make comparative statements on the quality of the melodies, adding harmonization and changing the instrumentation can influence the outcome of the experiment, even if the melodies remain the same.

Moreover, we are interested in the following question: does harmonization and instrumentation render the evaluation task of generated melodies more effective?

## 2 GENERATION OF MUSIC STIMULI

In this experiment we compare two different algorithms for melody generation: SuperWillow (van der Merwe and Schulze, 2010) and Melody-RNN (Waite et al., 2016), a melody generation algorithm from Google's open-source project Magenta[1].

The two algorithms have vastly different approaches to melody generation, one being based on Markov Chains, and the second on Long-Short Term Memory Neural Networks.

### 2.1 FF-Harmonizer

To test the assumption that varying the musical context of the melody will affect the evaluation, we designed a simple Harmonizer based on a Feed Forward Neural Network, trained on a lead sheet dataset (Simonetta et al., 2018), and applied it to the melodies generated by the systems described above. While for the scope of this experiment harmonization could have been manually added, we decided to use automatic Harmonization to better imitate the setting of computer-generated music.

FF-harmonizer takes an input melody of *T* bars and generates a corresponding chords sequence $Y = y_1, y_2,..., y_N$, in which *N* represents the length of the sequence. Each label is chosen from a set of 48 chords (we consider only triads in root position), specifically a triad chord is built for each degree of the chromatic scale (C, C\#, D, ..., B), and declined in the four possible qualities: major, minor, diminished, and augmented. During the training phase each musical piece of the corpus is encoded in a sequence of vectors $X = \vec{x}_1,\ldots,\vec{x}_1,\vec{x}_N$ representing the notes of the piece through a many-hot-encoding. The corresponding output is a sequence of numbers $Y = y_1, y_2,..., y_N$ representing the chords, each notes vector $\vec{x}_i$ corresponds to a chord $y_i$. In the prediction phase, the melody to be harmonized is represented by a sequence of vectors **X**, the length of the which depends on the harmonization frequency. For instance, considering a melody of **T** bars, if the user wants to generate a chord for each bar, then the length of the sequence of input vectors will be **T**, but if the user wants a chord every half bar, then there will be 2**T** input vectors, as the model predicts a chord for each of the input vectors.

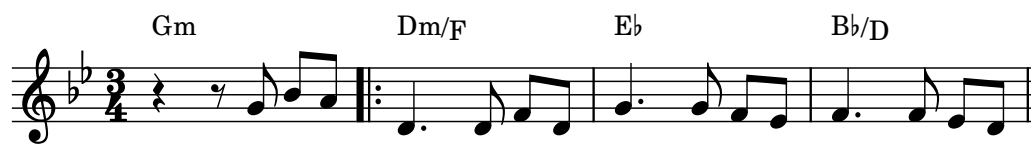

Figure 1. An example taken from a lead sheet from the dataset used to train the harmonization system.

---

[1] https://magenta.tensorflow.org

More specifically, consider the example song shown in Figure 1. Every file is loaded and parsed by the music21 library (Cuthbert and Ariza, 2010), which converts it to a specialized representation called a *Stream* object. Once the song is loaded, the following happens:

1. The algorithm finds the key of the piece and stores the tonic (i.e., G in the example);
2. The first chord is encoded, in this case G minor. This is represented by a number between 0 and 47, calculated by the sum of the distance $d$ in semitones between the tonic of the piece and the root of the chord (in this case $d = 0$) multiplied by 4, and the quality of the chord (major = +0, minor = +1, diminished = +2, augmented = +3). The result of this operation in the example is the label: 4d + *minor* = 1;

The algorithm then computes the vector of notes below the chord considered, in this example G, Bflat and A. Each note is encoded as the distance in semitones from the tonic of the piece (i.e., G): distance(tonic, G) → $d = 0$, distance(tonic, Bflat) → $d = 3$, distance(tonic, A) → $d = 2$. Then the vector of notes is filled using many-hot encoding: each vector position corresponding to values of $d$ is set to 1, all other elements are left to 0. The result of this operation in the example is: [1,0,1,1,0,0,0, 0, 0, 0, 0, 0].

The encoding result for this example is:

[1, 0, 1, 1, 0, 0, 0, 0, 0, 0, 0, 0] ↔ 1
[0, 0, 0, 0, 0, 0, 0, 1, 0, 0, 1, 0] ↔ 29
[1, 0, 0, 0, 0, 0, 0, 1, 0, 1, 0] ↔ 32
[0, 0, 0, 0, 0, 0, 0, 1, 1, 0, 1, 0] ↔ 12

The neural network is then trained to predict the chord given this one-hot encoding. When harmonizing a melody, a similar encoding will be produced for the given melody segments, and the network will predict the most likely chord.

## 2.2  Audio Rendering

Both the generation algorithms and the harmonizer only output music in symbolic format (either Midi or MusicXML). This means that, for the listening-based questionnaires we will describe below, it is necessary to render the generated files in score notation as audio files. Moreover, the harmonizer module outputs chords annotations, but not a complete arrangement defining how those chords should be played, so that this additional step towards the final rendering must be taken care of.

To render the monophonic, non-harmonized files, we simply used MuseScore's export feature to turn the score files into .mp3 files, using the default Grand Piano as the instrument playing the melody, which is arguably a popular default rendering choice among researchers (Carnovalini and Rodà, 2020).

The harmonized files needed an arrangement, as discussed above. The same arrangement was used for all the files, having the melody played simultaneously by a violin and a flute, a cello playing the root of the chord for the duration of each chord. A rhythmic section of two guitars was also added, which would play the selected chord in open position (following a chart of open-chords positions).

**Figure 2**. Two measures showing the output of the harmonization system, arranged for the "Group" modality used in the study.

One guitar would imitate fingerpicking by alternating low and high notes on the beats, and the other imitates strumming by playing all the notes of the chord following a typical folk rhythmic pattern. This instrument's volume was set to 40% while the other instruments were left at full volume. MuseScore export feature was again used for the generation of the .mp3 files, but in this case, it interprets the chords annotations by adding a piano playing the chord at the beginning of each measure, which was left as it helps immediately clearly establish the current harmony. Figure 2 shows one example of such arrangement over two measures harmonized by our algorithm.

## 3 EXPERIMENT

### 3.1 Subjects

Seventy-three responses were collected, of which twelve were discarded for failing the attention test or taking less than three minutes and a half to complete the survey (listening to all the melodic excerpts once takes about two minutes). The sixty-one remaining participants have an average age of 28.24 years (St. dev.: 10.85 years) and are divided in 41 males, 17 females, and three persons who did not specify their gender. Table 1 reports the responses to the musical experience question.

Table 1. Self-reported level of musical expertise of the participants.

| Level of Expertise | Count |
|---|---|
| 1. Nonmusician | 6 |
| 2. Music-loving nonmusician | 17 |
| 3. Amateur musician | 8 |
| 4. Serious amateur musician | 5 |
| 5. Semi-professional musician | 15 |
| 6. Professional Musician | 10 |

### 3.2   Method

Eight melodies were generated by the two systems, four melodies for each one, presented in Section 2. These eight original piano-only melodies were harmonized (group version) by following the above specified procedure. We will refer to these two versions of the same eight melodies as "modalities".

By means of a web interface, the users evaluate four melodies selected at random in one modality, and then the other four melodies in the other modality: the order of modalities is randomized (i.e., sometimes piano only comes first, other times the group version comes first) and the order of the melodies is randomized as well, again to avoid priming effects.

The users are asked to rank the first four musical excerpts based on their pleasantness/beauty, by dragging and dropping the html music player elements forming a leaderboard where the first element was the most pleasant/beautiful and the fourth was the least so.

The procedure repeats for the second set of four melodies, that use the modality that was not evaluated before. Note that in both modalities, the users evaluate two melodies for each algorithm.

At the end of the listening test, the users were asked to report their age and gender (but were offered the possibility not to respond) and to self-report their musical expertise by selecting the appropriate category among six (Zhang and Schubert, 2019). The participants had to agree to the informed consent before starting the test and had to respond to an attention check in the last page. Failing to do any of these two things resulted in automatic exclusion from the study. The time taken to respond to the questionnaire was also saved.

### 3.3   Results

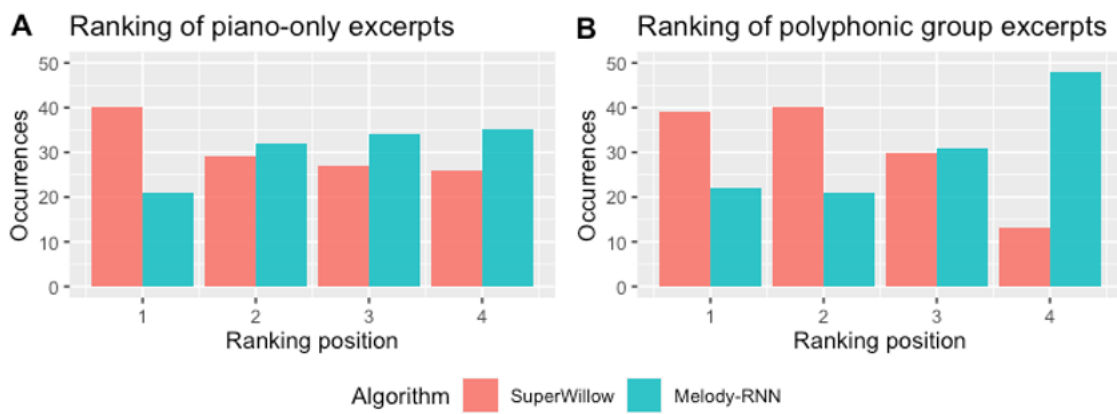

Figure 3. The results of the listening test, in aggregated form. A shows results for the piano-only excerpts, whereas B the results for the harmonized ones. They report the positional rankings, where the lower number means higher position in the ranking and thus more pleasing excerpt. Red bars are relative to SuperWillow excerpts, while blue bars to Melody-RNN.

Figure 3 shows the distributions of rankings for the two algorithms under different modalities. In all cases, the distributions are statistically different with 95% confidence according to a Mann-Whitney U test. In both the modalities the excerpts generated with SuperWillow were preferred by the subjects, i.e., the melodies generated by SuperWillow obtained globally a better (lower rank) position. More interesting, as Figure 3 shows, the difference between the algorithms is more evident when using the group modality. In the following section we try to assess whether the increase in difference is statistically significant.

To do so, a prediction model that estimates the evaluation based on the algorithm, the modality, and possibly the interaction with other features needs to be constructed. Since we consider the predicted value as ordinal rather than numerical, we used Cumulative Link Models (CLM), and used Analysis of Variance (ANOVA) to determine the importance of different factors.

The models we constructed try to predict the ranking of a piece based on the interaction between algorithm used, modality used, and musicianship of the evaluator. For this feature, the self-assessed level of musicianship was reduced to a binary value, distinguishing nonmusicians from musicians. The model was written as follows in R, using the function `clm` from the `ordinal` package:

```
model = clm(ranking ~ algorithm*modality*musicianship)
```

The ANOVA underlines that the used algorithm and the interaction between algorithm and modality are strongly significant ($p \ll 0.001$). Moreover, the musicianship is also strongly significant ($p \ll 0.001$).

## 4  DISCUSSION

While the first result is that SuperWillow was preferred to Melody-RNN in our limited test set, that was not the goal of this work, and this result is not very significant since many other parameters would need to be considered to properly compare the two algorithms.

The more interesting results coming from the ANOVA analysis indicate that, while the difference between the algorithms is the main predictor (as could be expected from the difference in ratings of the two algorithms), the modality had indeed an effect on the evaluation, and that this effect varied depending on the algorithm. Musicianship is also impactful, as expected from scientific literature (Amabile, 1983).

Figure 4 shows the effect of the interaction between modality and algorithm on the ranking. We can see that the SuperWillow melodies are more likely to be in first position when harmonized, and the harmonized Melody-RNN excerpts are more likely to be in last position.

This suggests that the harmonization have not changed the global listeners' preference (SuperWillow was preferred in both the modalities), instead it rendered more evident this preference. If this result would be confirmed by further experiments (with more generation systems and musical pieces), it could imply a positive effect of harmonization and instrumentation in the evaluation process, given it contributes to obtain more clear results.

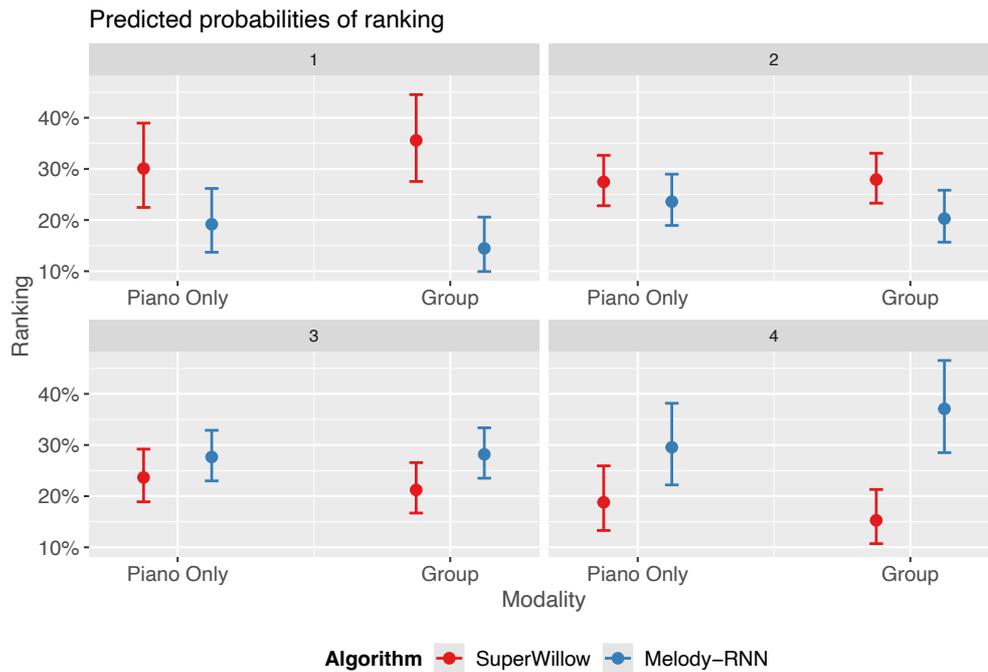

**Figure 4**. Interaction of chosen algorithm and modality. For each of the possible rankings (1 through 4) the probability of receiving that ranking is reported for both algorithms and modalities. The probability of receiving ranking 1 (lower ranking is better) grows with the harmonization for SuperWillow, but diminishes for Melody-RNN, and the opposite is true for ranking 4.

Of course, this effect could be also explained by the idea that certain melodies benefit from the harmonization process, while others are not really affected. One motivation could be that SuperWillow melodies sometimes include long notes, which can be dull-sounding when played monophonically on the piano, an instrument with a limited sustain, but are better perceived when played on sustain-heavy instruments like violin and flute along with an accompaniment that keeps the rhythm.

Obviously, this observation only applies to our example melodies, and different algorithms or different melodies could show different effects. Still, this observation supports the hypothesis that harmonization influences how melodies are perceived and evaluated.

## 5    CONCLUSIONS

In this paper, we designed an evaluation-by-comparison study to test whether harmonization and instrumentation can have an impact on the evaluation of computer-generated music, often used in the sonification of mixed reality systems. The study collected sixty-one responses from participants who evaluated melodies generated by two different algorithms for music generation which were presented both rendered as played only by a piano mono-phonically or by a small polyphonic group.

The results show that while the general comparison of the two algorithms was not changed by the presence of harmonization, it still influenced how the participants responded, affecting in different ways the evaluation of the melodies generated by the two algorithms. We observed that harmonization does not change the global trend of the preferences, but it makes the judgements clearer, supporting the conclusion that adding harmonization would be desirable when the evaluation of computer-generated melodies is required.

The main limitation of this study is the few systems (two) and melodies (eight in total) considered, numbers which were kept low due to the questionnaire fatigue, which was not negligible for a web-based survey as testified by the amount of failed attention checks. Nevertheless, it clearly shows the influence that the musical context (harmony and arrangement) can have on the evaluation of melodies. Researchers should be aware that such an effect exists and use caution when designing their evaluation methods.

## ACKNOWLEDGMENTS

The first author is funded by University of Padova. Many thanks go to Fiorella Del Popolo Cristaldi, who helped design the statistical analyses, as well as all the folks at CSC and VUB CC lab for their helpful comments on this work.